\documentclass[12pt]{article}
\usepackage{amssymb}
\begin{document}

\def\kbar{{\mathchar'26\mkern-9muk}}  
\def\bra#1{\langle #1 \vert}
\def\ket#1{\vert #1 \rangle}
\def\vev#1{\langle #1 \rangle}
\def\tr{\mbox{Tr}\,}
\def\ad{\mbox{ad}\,}
\def\ker{\mbox{Ker}\,}
\def\im{\mbox{Im}\,}
\def\m@th{\mathsurround=0pt}
\def\eqalign#1{\null\,\vcenter{\openup 3pt \m@th
\ialign{\strut\hfil$\displaystyle{##}$&$\displaystyle{{}##}$\hfil
\crcr#1\crcr}}\,}

\title{Topology at the Planck Length}

\author{J. Madore, \  L.A. Saeger \\
Laboratoire de Physique Th\'eorique et Hautes 
Energies\thanks{Laboratoire associ\'e au CNRS, URA D0063}\\
Universit\'e de Paris-Sud, B\^at. 211, F-91405 Orsay
}

\date{May, 1997}

\maketitle

\abstract{A basic arbitrariness in the determination of the topology of
a manifold at the Planck length is discussed. An explicit example is
given of a `smooth' change in topology from the 2-sphere to the 2-torus
through a sequence of noncommuting geometries. Applications are
considered to the theory of $D$-branes within the context of the
proposed $M$(atrix) theory.}

\vfill
\noindent
PACS 02.40.-k, 04.20.Cv\\
\noindent
LPTHE Orsay 97/34
\medskip
\eject

\parskip 4pt plus2pt minus2pt

\section{Motivation and notation}

Since the early efforts of Wheeler~\cite{Whe57} in this direction it has
often been speculated that the topology of space might not be a well
defined dynamical invariant. There can of course be no smooth time
evolution of a space of one topology into that of another; a classical
space cannot change topology without the formation of a singularity.
However the `true' description of space and of space-time should
reasonably include quantum fluctuations and it is possible that a
quantum space-time exists which seems in a quasi-classical approximation
to evolve from a space of one topology to that of another, the `exact'
quantum space-time being a sum over many topologies.

Noncommutative geometry furnishes a possible alternative mathematical
language in which one can also discuss this question. A change in
topology is possible simply because even classically the topology of a
manifold is not a well defined quantity at all length scales. A change
in space topology can occur if the dynamical evolution of space-time is
such that the space enters a regime in which its description requires
the use of noncommutative geometry. This could be expected to occur near
a classical singularity. We are able to treat this problem only in 2+1
dimensions where we can identify space with a smooth compact surface of
genus~$g$.  Even here we can give no reasonable field equations which
would dynamically implement the change of space topology which we
consider.

In Section~2 we propose a definition of a fuzzy surface of ill-defined
topology. In Section~3 we discuss the differential structures of fuzzy
surfaces and we describe in detail an explicit example of topology
change from the 2-sphere to the 2-torus through a sequence of such fuzzy
surfaces.  We speculate on the analogous transition between two compact
surfaces of arbitrary genus. Finally in Section~5 we discuss our results
in light of the recent fuzzy description of $D$-branes. 

When using in the noncommutative context a word which is usually defined
only for an ordinary differential manifold we enclose it in quotation
marks if there is a chance of ambiguity.

\section{Topological fuzzy surfaces}

The general definition of a fuzzy surface has been given
elsewhere~\cite{Mad95, Mad97}. Points are replaced by elementary cells
(Planck cells) of a quantum of area. If the `surface' is to be in some
sense compact then there can only be a finite number of cells and the
structure algebra must be of finite dimension. It is usually taken to be
the algebra $M_n$ of $n\times n$ complex matrices. The `topology' is
encoded in a filtration ${\cal P}_h$ of the algebra which we shall
introduce for each genus $h$. The differential structure is encoded in
the differential calculus over the algebra.  We shall suppose that there
exist fuzzy versions of compact surfaces $\Sigma_h$ of arbitrary genus
$h$ although we know of explicit constructions only in the particular
cases $h = 0$ and $h = 1$.

Let $f$ be an element of $M_n$. For each $h$ we shall introduce a norm
$\Vert f \Vert^2_{h,n}$ If the sequence $\Vert f \Vert_{h,n}$ has a
limit for some value of $h$ then we consider $f$ to be a matrix
approximation to a function $\tilde f$ on $\Sigma_h$.  The limiting
procedure is rather obscure and we shall consider it only on the algebra
of polynomials in a set of basic matrices, the `coordinates'. The choice
of this set will define the filtration and hence the value of $h$.
Since $M_n$ is a simple \hbox{$*$-algebra} a morphism is necessarily of
the form $f \mapsto u^{-1} f u$ for some hermitian $u$.  In the
commutative limit these maps tend to symplectomorphisms of $\Sigma_h$
into itself~\cite{deWHopNic88, Hop89, FaiFleZac89, FloIliTik89,
BorHopSchSch91, Mad92}.  A general `coordinate transformation' would be
a map of the form $f \mapsto \phi(f)$ which respects the algebraic
structure only in the commutative limit. If such a map is singular in
the commutative limit then the resulting transformation is a change of
topology.

\subsection{Genus zero}

Let $\tilde x^a$ be the cartesian coordinates of ${\mathbb R}^3$ and
$g_{ab}$ the euclidean metric.  The ordinary round sphere $S^2$ of
radius $r$ is defined by the constraint
$g_{ab}\tilde x^a \tilde x^b - r^2 = 0$.  The algebra ${\cal C}(S^2)$ of
smooth functions on $S^2$ is a completion of the quotient of the algebra
of polynomials in the $\tilde x^a$ by the ideal generated by those which
contain $g_{ab}\tilde x^a \tilde x^b - r^2$ as a factor.

A fuzzy version of the sphere~\cite{Ber75, deWHopNic88, Hop89,
FaiFleZac89, FloIliTik89, BorHopSchSch91, Mad95} is constructed using an
$n$-dimensional irreducible representation of the Lie algebra of the
group $SU_2$.  We let $x^a$, for $1 \leq a \leq 3$, be the generators
and we raise and lower indices using the Killing metric $g_{ab}$. We
introduce a macroscopic length scale $r$, the radius of the sphere, and
a microscopic area scale $\kbar$ which are related, for large $n$, by
the equation
$$
{4\pi r^2 \over 2\pi\kbar} = n.                                  \eqno(2.1.1)
$$
The integer $n$ counts the number of elementary cells of area
$2\pi\kbar$.  The Casimir relation is written as $g_{ab} x^a x^b = r^2$
and the commutation relations of the `coordinates' $x^a$ are given by
$$
[x_a, x_b] = i \kbar C^c{}_{ab} x_c, \qquad 
C_{abc} = r^{-1} \epsilon_{abc}.                                 \eqno(2.1.2)
$$
We shall consider the length scale $r$ as fixed and so 
$\kbar \rightarrow 0$ in the limit as $n \rightarrow \infty$.
We can identify therefore
$$
\lim_{n\rightarrow\infty} x^a = \tilde x^a.                      \eqno(2.1.3)
$$

Any matrix $f$ can be written as a polynomial $f(x^a)$
in the $x^a$,
$$
f = \sum_0^l {1\over k!} f_{a_1 \cdots a_k} x^{a_1} \cdots x^{a_k},
                                                                 \eqno(2.1.4)
$$
where the $f_{a_1 \cdots a_k}$ are completely symmetric and trace-free.
We can associate to $f$ the function $\tilde f = f(\tilde x^a)$. Set
$\tilde f = \phi_n(f)$. We have defined then a vector-space map
$$
M_n \buildrel \phi_n \over \longrightarrow \tilde {\cal P}(S^2).
$$
This map cannot of course respect the product
structures of the respective algebras but if $f$ and $g$ are two
polynomials of order less than some integer $l$ then one can show
that
$$
\phi_n(fg) - \phi_n(f) \phi_n(g) = o(l/n).                        \eqno(2.1.5)
$$
For each integer $0 \leq l \leq n-1$ introduce the vector space 
${\cal P}_{0,l}$ of symmetric polynomials of order $l$ in the $x^a$. 
Obviously
$$
{\cal P}_{0,l} \subset {\cal P}_{0,l+1}, \qquad 
\bigcup_{l=0}^{n-1} {\cal P}_{0,l} = M_n.                         \eqno(2.1.6)
$$
The filtration of the algebra $M_n$ which defines the sphere is given by
the ${\cal P}_{0,l}$ and $\phi({\cal P}_{0,l})$ is a filtration of
the polynomials of order $n$ on the sphere.

We define the norm $\Vert f \Vert_n$ of an element $f \in M_n$ as
$$
\Vert f \Vert^2_n = {1\over n} \tr (f^* f).                       \eqno(2.1.7)
$$
In particular we find that
$$
\Vert x^a \Vert^2_n = {1\over 3} r^2.                             \eqno(2.1.8)
$$
We introduce the norm of an element $\tilde f \in {\cal C}(S^2)$ as
$$
\Vert \tilde f \Vert^2 = 
{1 \over 4 \pi r^2} \int_{S^2} \vert \tilde f \vert^2.
$$
Then if $f \in {\cal P}_{0,l}$ we have
$$
\Vert f \Vert^2_n - \Vert \tilde f \Vert^2 = o(l/n).              \eqno(2.1.9)
$$
The norm of a generic element of ${\cal P}_{0,l}$ grows as $l$.

\subsection{Genus one}

Let $r$ be again a length scale and consider the torus ${\mathbb T}^2$
defined to be the subset of ${\mathbb R}^2$ with coordinates 
$(\tilde x, \tilde y)$ subject to the conditions 
$0 \leq \tilde x, \tilde y \leq 2\pi r$. Consider the two functions
$$
\tilde u = e^{i \tilde x/r}, \qquad \tilde v = e^{i \tilde y/r}.  \eqno(2.2.1)
$$
The algebra ${\cal C}({\mathbb T}^2)$ of smooth functions on 
${\mathbb T}^2$ is a completion of the algebra of polynomials in 
$\tilde u$ and $\tilde v$.

A fuzzy version of the torus was constructed by Weyl~\cite{Wey31}, 
Schwinger~\cite{Sch60} and others~\cite{Flo89, Ald93, AthFloNic96} 
to describe a finite version of quantum mechanics. One
introduces elements $u$ and $v$ which satisfy the Weyl relation
$$
uv = q vu                                                         \eqno(2.2.2)
$$
as well as the constraints 
$$
u^n = 1, \qquad v^n = 1, \qquad q = e^{2 \pi i /n}.               \eqno(2.2.3)
$$
The algebra generated by $u$ and $v$ is isomorphic then to the matrix
algebra $M_n$. Define the area parameter $\kbar$ by the relation
$$
{(2\pi r)^2 \over 2\pi\kbar} = n.                                 \eqno(2.2.4)
$$
This is the same as (2.1.1) if one replaces the area $4 \pi r^2$ of the
sphere of radius $r$ by the area $(2 \pi r)^2$ of the torus.  It is
worth noticing that when described by the algebra $M_n$ the torus has
$n$ `cells'; each observable can take $n$ possible values. It is
therefore to be compared with an approximation on a lattice with
$$
{2 \pi r \over \sqrt n} = \sqrt{2\pi\kbar}
$$
as unit of length.

An explicit form for $u$ and $v$ can be easily found~\cite{Sch60}.
There is an orthonormal basis $\ket{j}_1$, $0 \leq j \leq n-1$, of 
${\mathbb C}^n$ such that $u$ and $v$ are given by
$$
v \ket{j}_1 = \ket{j+1}_1, \qquad u \ket{j}_1 = q^j \ket{j}_1,   \eqno(2.2.5)
$$
and such that the cyclicity condition
$$
v \ket{n-1}_1 = \ket{0}_1                                        \eqno(2.2.6)
$$
holds. One can introduce matrices $x$ and $y$ defined by the relations
$$
u = e^{ix/r}, \qquad v = e^{iy/r}.                               \eqno(2.2.7)
$$
In the basis $\ket{j}_1$ it is obvious that one can choose $x$ such that
$$
x \ket{j}_1 = {\kbar \over r} j \ket{j}_1.                       \eqno(2.2.8)
$$

There is also an orthonormal basis $\ket{j}_2$ in which the $v$ is 
diagonal.  The two bases are related by the `Fourier
transformation'~\cite{Sch60}
$$
\ket{l}_2 = {1\over \sqrt n} \sum_{j=0}^{n-1} q^{-jl} \ket{j}_1, \qquad
\ket{j}_1 = {1\over \sqrt n} \sum_{l=0}^{n-1} q^{+jl} \ket{l}_2. \eqno(2.2.9)
$$
If we define, for arbitrary $z \in {\mathbb C}$,
$$
f(z) = {1\over n} \sum_{l=0}^{n-1} q^{-zl} = 
{1\over n} {q^{-zn} - 1 \over q^{-z} - 1}                       \eqno(2.2.10)
$$
then we can write 
$$
\phantom{}_1 \vev{j^\prime \vert j}_1 = f(j^\prime - j).        \eqno(2.2.11)
$$
The Fourier transformation is unitary because of the relations 
$f(0) = 1$ and $f(j) = 0$, $j \neq 0$. A short calculation yields the
relation
$$
\phantom{}_1 \bra{j^\prime} [x,y] \ket{j}_1 = 
i \kbar (j^\prime - j) f^\prime(j^\prime - j).                  \eqno(2.2.12)
$$
In the $n \rightarrow \infty$ limit we must have
$f(z) \rightarrow \delta(z)$ and therefore 
$z f^\prime \rightarrow - \delta$. We recover then the commutation 
relation
$$
[x,y] = - i \kbar                                               \eqno(2.2.13)
$$
which is equivalent to the Weyl relation (2.2.2).

Because $q \rightarrow 1$ as $n \rightarrow \infty$ we can identify
$$
\lim_{n\rightarrow\infty} u = \tilde u, \qquad 
\lim_{n\rightarrow\infty} v = \tilde v.
$$
Introduce $u^\alpha = (u,v)$. Any matrix $f$ can be written as a
polynomial $f(u^\alpha)$ in the $u^\alpha$:
$$
f = \sum_0^l {1\over j!} f_{\alpha_1 \cdots \alpha_j} 
u^{\alpha_1} \cdots u^{\alpha_j}                                \eqno(2.2.14)
$$
where the $f_{\alpha_1 \cdots \alpha_j}$ are completely symmetric.  We can
associate to $f$ the function $\tilde f = f(\tilde u^\alpha)$. Set 
$\tilde f = \phi_n(f)$. We have defined then a vector-space map
$$
M_n \buildrel \phi_n \over \longrightarrow \tilde {\cal P}({\mathbb T}^2).
$$
As above, if $f$ and $g$ are two polynomials of order less than some
integer $l$ then one can show that
$$
\phi_n(fg) - \phi_n(f) \phi_n(g) = o(l/n).                       \eqno(2.2.15)
$$
For each integer $0 \leq l \leq n-1$ introduce the vector space 
${\cal P}_{1,l}$ of symmetric polynomials of order $l$ in the $u^\alpha$. 
Obviously
$$
{\cal P}_{1,l} \subset {\cal P}_{1,l+1}, \qquad 
\bigcup_{l=0}^{n-1} {\cal P}_{1,l} = M_n.                        \eqno(2.2.16)
$$
The filtration of the algebra $M_n$ which defines the torus is given by
the ${\cal P}_{1,l}$ and $\phi_n ({\cal P}_{1,l})$ is a filtration of
the polynomials of order $n$ on the torus.

We define again the norm $\Vert f \Vert_n$ of an element 
$f \in M_n$ by (2.1.7). In particular we find that
$$
\Vert u^\alpha \Vert^2_n = 1.                                    \eqno(2.2.17)
$$
We define the norm of an element $\tilde f \in {\cal C}({\mathbb T}^2)$ 
as
$$
\Vert \tilde f \Vert^2 = 
{1 \over (2 \pi r)^2} \int_{{\mathbb T}^2} \vert \tilde f \vert^2.
$$
Then if $f \in {\cal P}_{1,l}$ we have
$$
\Vert f \Vert^2_n - \Vert \tilde f \Vert^2 = o(l/n).             \eqno(2.2.18)
$$
The norm of a generic element of ${\cal P}_{1,l}$ grows as $l$.

\subsection{Higher genera}

We conjecture that the construction of Section~2.1 and Section~2.2 can
be extended to arbitrary genus. Let $\Sigma_h$ be a surface of genus $h$
and choose generators $x^i$ of $M_n$ which define in the limit 
$n \rightarrow \infty$ coordinates $\tilde x^i$ on $\Sigma_h$. There
might be a large number $d$ of the $x^i$ which satisfy $d-2$ relations.
For each integer $0 \leq l \leq n-1$ introduce the vector space 
${\cal P}_{h,l}$ of symmetric polynomials of order $l$ in the $x^i$ such
that
$$
{\cal P}_{h,l} \subset {\cal P}_{h,l+1}, \qquad 
\bigcup_0^{n-1} {\cal P}_{h,l} = M_n.                            \eqno(2.3.1)
$$
The filtration of the algebra $M_n$ which defines $\Sigma_h$ is given
then by the ${\cal P}_{h,l}$.  

We define again the norm $\Vert f \Vert_n$ of an element $f \in M_n$ as
(2.1.7).  We introduce the norm of an element 
$\tilde f \in {\cal C}(\Sigma_h)$ as
$$
\Vert \tilde f \Vert^2 = 
{1 \over \hbox{Vol}(\Sigma_h)} \int_{\Sigma_h} \vert \tilde f \vert^2
$$
Then, if $f \in {\cal P}_{h,l}$ we should have
$$
\Vert f \Vert^2_n - \Vert \tilde f \Vert^2 = o(l/n).             \eqno(2.3.2)
$$
The norm of a generic element of ${\cal P}_{h,l}$ grows as $l$.  We
shall return to this in Section~3.3.

\subsection{Continuous transitions}

From the point of view of noncommutative geometry a transition is
possible between space-times of different topology simply because an
individual space-time is never completely in a `pure' topological state.
As long as $\kbar$ is not equal to zero the correct description of every
surface is given in terms of a filtration of the matrix algebra $M_n$
for some (very large) integer $n$. A transition occurs when one
filtration becomes more appropriate than another. Below we shall
introduce differential calculi on $M_n$ and we shall be in a position to
speak of the noncommutative analog of a smooth scalar field.  A
generic such field $\tilde f$ on a surface $\Sigma_h$ of genus $h$ must
have finite action $\tilde S_h(\tilde f)$ and every other action $\tilde
S_{h^\prime}(\tilde f)$ must be `almost always' infinite. If during the
time evolution the action changes so that $\tilde f$ has finite
action for the genus $h^\prime \neq h$ then this means that the surface
has evolved towards a different topology.

The difference in topology between the sphere and the torus is expressed
in a discontinuity in the functions 
$\tilde x^a = \tilde x^a(\tilde u^\alpha)$. These discontinuities will
not show up in the norm (2.1.7) we have put on $M_n$. They do show up
however if we use the action as norm since it contains derivatives. We
shall discuss in the following section how a topological transition can
be induced using the partition function after we have introduced
differential calculi and the associated scalar-field actions.

\section{Smooth fuzzy surfaces}

Every surface can of course be endowed with a differential structure and
the associated de~Rham calculus of differential forms. To speak of a
smooth fuzzy surface we must be able to define a differential calculus
on each fuzzy $\Sigma_h$ which in some sense has the de~Rham calculus as
a limit. Since the de~Rham calculus is based on the derivations of the
algebra of functions it is natural to require that for each $h$ the
differential calculus over $M_n$ be based on derivations. This idea was
first suggested by Dubois-Violette~\cite{Dub88} and developed by
Dubois-Violette {\it et al.}~\cite{DubKerMad90}. We shall use a modified
version proposed later by Dimakis~\cite{DimMad96}. Since we shall
restrict our considerations here to scalar fields and shall not
therefore need explicitly the differential calculus we shall not enter
into the details of its construction. Some more details will be given
where necessary in Section~4.  We recall that a classical scalar field
defined on a fuzzy surface of any genus is an element of $M_n$. The form
of the matrix determines the genus of the surface on which it is to be
considered an approximation to a regular function.

For each $h$ we shall define a differential structure over $M_n$ in
order to be able to speak of the noncommutative analogue of a smooth
scalar field. We can then define an action $S_{h,n}$ which tends to the
action $\tilde S_h$ of a complex classical field on $\Sigma_h$. Let $f$
be an element $f$ of $M_n$ which tends to a function $\tilde f$ on
$\Sigma_h$.  Then we have
$$
\lim_{n \rightarrow \infty} S_{h,n}(f) =  \tilde S_h(\tilde f).
$$
We shall use the action to define a Sobolov-like norm on the matrices
and a Sobolov norm on the limit functions. We shall return to this in
Section~3.4.

\subsection{Genus zero}

The derivations 
$$
e_a = {1 \over i\kbar} \ad x_a                                   \eqno(3.1.1) 
$$
satisfy the commutation relations
$$
[e_a, e_b] = i \kbar C^c{}_{ab} e_c.                            \eqno(3.1.2)
$$
In the commutative limit these derivations tend towards vector fields 
$\tilde e_a$ on the sphere defined by the action of the Lie algebra of
$SO_3$.  The relation $\tilde x^a \tilde e_a = 0$ defines the space 
${\cal X}_0$ of vector fields on the sphere as a (projective) submodule
of the free ${\cal C}(S^2)$-module generated by the $\tilde e_a$. 

We choose the differential calculus $\Omega^*(M_n)$ defined in
terms of the $e_a$, that is, with the 1-forms defined by the relation
$$
df (e_a) = e_a f.                                                \eqno(3.1.3)
$$
Since the sphere $S^2$ is not parallelizable the differential calculus
must be defined on a parallelizable bundle over it. The details of
this have been described elsewhere~\cite{Mad92, GroKliPre97b, CarWat97}.
It is important only to recall the existence of a special basis or frame
$\theta^a$ which is dual to the derivations and which commutes with the
elements of the algebra.

We define the action $S_{0,n}(f)$ of the matrix $f$ on the surface
$\Sigma_0$ as the trace
$$
S_{0,n}(f) = 
{1 \over n} \tr (f^* (\Delta_0 f + \mu^2) f + V(f^*f))            \eqno(3.1.4) 
$$
where the laplacian is the covariant laplacian with respect to the
geometry we have put on the sphere and $V(f^*f)$ is an arbitrary
(positive) potential function. The normalization has been chosen so that
$$
\lim_{n \rightarrow \infty} S_{0,n}(f) =  \tilde S_0(\tilde f)    \eqno(3.1.5)
$$
where $S_0(\tilde f)$ is the usual action of the classical complex 
scalar field $\tilde f$. Obviously we shall have
$$
S_{0,n}(f) = 0(n)
$$
for almost all elements $f \in M_n$. 

It is of interest to note that because of the identity
$$
df^* * df = {1\over 2} e_af^* e^af \theta^1 \theta^2 \theta^3     \eqno(3.1.6)
$$
it is possible to write the action (3.1.4) without explicitly using the
derivations. The 2-form $*df$ is defined using a straightforward
generalization of the standard duality in forms which relies on the
existence of the preferred frame.

\subsection{Genus one}

The vector fields
$$
\tilde e_1 = \partial_{\tilde x}, \qquad 
\tilde e_2 = \partial_{\tilde y}
$$
form a basis of the free ${\cal C}({\mathbb T}^2)$-module ${\cal X}_1$
of vector fields on the torus. Their action on the generators is given by
$$
\begin{array}{ll}
\tilde e_1 \tilde u = i r^{-1} \tilde u, &\tilde e_1 \tilde v = 0, \\
\tilde e_2 \tilde v = i r^{-1} \tilde v, &\tilde e_2 \tilde u = 0.  
\end{array}                                                      \eqno(3.2.1)
$$
and of course they commute:
$$
[\tilde e_1, \tilde e_2] = 0.
$$

The dual de~Rham 1-forms $\tilde \theta^\alpha$ are given by
$$
\tilde \theta^1 = d \tilde x = - i r \tilde u^{-1}  d \tilde u,  \qquad
\tilde \theta^2 = d \tilde y = - i r \tilde v^{-1}  d \tilde v.  \eqno(3.2.2) 
$$
Because the torus does not have as large an invariance group as the
sphere it is more difficult to find a differential calculus over $M_n$
which tends to the de~Rham calculus. This fact leads us to believe that
the introduction of appropriate noncommutative differential calculi over
fuzzy surfaces of higher genera will be a delicate matter.

Were it not for the extra constraints (2.2.3) which distinguish the
`quantum' torus from the `quantum' plane we could have used the
`quantum' analog of (3.2.1) and introduced a differential calculus based
on the outer derivations $\delta_\alpha$ defined by
$$
\begin{array}{ll}
\delta_1 u = i r^{-1} u, &\delta_1 v = 0, \\
\delta_2 v = i r^{-1} v, &\delta_2 u = 0.  
\end{array}                                                      \eqno(3.2.3)
$$
If we extend formally the algebra and admit hermitian elements
$x = - i r \log u$ and $y = - i r \log v$ then these derivations become
inner and can be written, using the relation (2.2.4) as
$$
e_1 =   {1 \over i\kbar} \ad y, \qquad
e_2 = - {1 \over i\kbar} \ad x.                                  \eqno(3.2.4)
$$
The associated frame is formally identical to (3.2.2):
$$
\theta^1 = - ir u^{-1} du, \qquad
\theta^2 = - ir v^{-1} dv.                                       \eqno(3.2.5)
$$
It is easy to see~\cite{DimMad96} that the associated differential
calculus admits a flat metric-compatible torsion-free linear connection.

But the above derivations $\delta_\alpha$ are not compatible with the
constraints (2.2.3).  With these constraints the algebra is a matrix
algebra and all derivations must be inner. This leads to problems. It is
of course in itself not surprising to encounter a situation where
`quantization' is inconsistent with certain constraints; this feature of
quantum mechanics was known to Dirac. Using the representations of
Section~2.2 the commutation relations
$$
[x,v] =   {\kbar \over r} v (1 - n P_2), \qquad
[y,u] = - {\kbar \over r} u (1 - n P_1)                           \eqno(3.2.6) 
$$
are easily derived. We have here introduced the projectors
$$
P_2 = \ket{n-1}_1 \bra{n-1}, \qquad P_1 = \ket{0}_2 \bra{0}.      \eqno(3.2.7)
$$
As every element of the algebra they can be expressed as polynomials in
the generators:
$$
P_2 = {1 \over n} \sum_0^{n-1} q^l u^l,\qquad
P_1 = {1 \over n} \sum_0^{n-1} v^l.                               \eqno(3.2.8)
$$
It follows that the action of the derivations (3.2.4) on the generators
of the algebra is given by
$$
\begin{array}{ll}
e_1 u = i r^{-1} u (1 - n P_2), &e_1 v = 0, \\
e_2 v = i r^{-1} v (1 - n P_1), &e_2 u = 0.
\end{array}                                                       \eqno(3.2.9)
$$
The highly singular projector term on the right-hand side of each of
these equations is due to the constraints (2.2.3). It is because of these
terms that we find $e_1 u^n = 0$ and $e_2 v^n = 0$ as we must.

The 1-forms dual to the derivations (3.2.4) are given by
$$
\theta^1 = - ir (1 - {n \over n-1} P_1) u^{-1} du, \qquad
\theta^2 = - ir (1 - {n \over n-1} P_2) v^{-1} dv.               \eqno(3.2.10)
$$
If we compare (3.2.10) with (3.2.2) we see that the $\theta^\alpha$
could in a weak way be considered to tend to the $\tilde\theta^\alpha$.
The problem of the singular limit of the differential calculus is hidden
however in the differentials $dP_\alpha$; the differential calculus
based on the derivations (3.2.4) does not tend to the de~Rham
differential calculus on the torus.

It was of course not necessary to use a differential calculus based on
derivations and one can introduce many another differential calculi over
the `quantum' torus.  There are in fact many which can be
constructed~\cite{DimMad96} based on derivations but which are not real.
It is easy to see however that whatever the definition of $du$ and $dv$
the 1-forms (3.2.5) cannot commute with the elements of the algebra and
that the resulting differential calculus will not have them as a
preferred frame. Also to define the action we will have to be able to
define a Laplace operator using the derivations.

To construct the torus we identified the points $\tilde x + 2\pi r$ with
$\tilde x$ and $\tilde y + 2\pi r$ with $\tilde y$. In the `quantized'
version this becomes the cyclicity condition (2.2.6) which gives rise to
the singular projector terms in the derivations (3.2.9).  One can
eliminate them by a procedure which is equivalent to folding, so to
speak, the torus at $\tilde x = \pi r$ or $\tilde y = \pi r$. For this
we suppose that $n = 2m$ is even in the formulae of Section~2.2 and we
consider the possibility of a differential calculus based on the
derivations of the form (3.2.4) with $x$ and $y$ replaced respectively
by $x^\prime$ and $y^\prime$ defined by
$$
x \ket{j}_1 = {\kbar \over r} (j + n F_j) \ket{j}_1,\qquad
y \ket{j}_2 = {\kbar \over r} (j + n G_j) \ket{j}_2.           \eqno(3.2.11)
$$
We shall suppose that $F_j, G_j \in {\mathbb Z}$ so that we have
$$
u = e^{ix^\prime/r}, \qquad v = e^{iy^\prime/r}.
$$
The matrices $x$ and $y$ are not defined then uniquely by the
Formulae~(2.2.7). This fact is related to the fact that only by using
additional topological conditions was von~Neumann able to deduce the
uniqueness of the representation of the Heisenberg commutation
relations. For a discussion of this and an introduction to the problems
connected with the quantization of the torus as a classical phase space
as well as reference to the previous literature on the subject we refer
to the lecture by Emch~\cite{Emc95} or to the recent article by
Narnhofer~\cite{Nar97}.

If we choose
$$
F_j = - \vert m - j \vert, \qquad 
G_j = \vert m - j - 1 \vert, \qquad n = 2m                      \eqno(3.2.12)
$$
and introduce the `step functions'
$$
\epsilon_1 \ket{j}_2 = \cases{-\ket{j}_2, \quad j \leq m-1, \cr
                              +\ket{j}_2, \quad j \geq m,}  \qquad
\epsilon_2 \ket{j}_1 = \cases{+\ket{j}_1, \quad j \leq m-1, \cr
                              -\ket{j}_1, \quad j \geq m}       \eqno(3.2.13)
$$
we find
$$
[x^\prime,v] =   {\kbar \over r} v (1 + n \epsilon_2 (1+P_2)), \qquad
[y^\prime,u] = - {\kbar \over r} u (1 + n \epsilon_1 (1+P_1)).  \eqno(3.2.14) 
$$

The commutation relations (3.2.14) are almost as singular as (3.2.6).
The presence however of the extra factors $\epsilon_\alpha$ permits us  
to `renormalize' the $e_\alpha$ and define
$$
e_1 =   {1\over n} {1 \over i\kbar} \ad y^\prime, \qquad
e_2 = - {1\over n} {1 \over i\kbar} \ad x^\prime.               \eqno(3.2.15)
$$
We find then in the $n\rightarrow\infty$ limit
$$
\begin{array}{ll}
e_1 u = i r^{-1} u \epsilon_1 (1+P_1), &e_1 v = 0, \\
e_2 v = i r^{-1} v \epsilon_2 (1+P_2), &e_2 u = 0.
\end{array}                                                      \eqno(3.2.16)
$$
We introduce the step functions
$$
\tilde\epsilon_1 = \cases{-1, \quad \tilde x < \pi r, \cr
                          +1, \quad \tilde x > \pi r,} \qquad
\tilde\epsilon_2 = \cases{+1, \quad \tilde y < \pi r, \cr
                          -1, \quad \tilde y > \pi r.}
$$
We can claim then that in a weak way
$$
\lim_{n\rightarrow\infty} \epsilon_\alpha (1+P_\alpha) = 
\tilde\epsilon_\alpha
$$
and comparing (3.2.1) with (3.2.16) we find that
$$
\lim_{n\rightarrow\infty} e_\alpha =
\tilde\epsilon_\alpha \tilde e_\alpha.                           \eqno(3.2.17)
$$
The limit of the derivations $e_\alpha$ are vector fields on the torus
which form a basis of ${\cal X}_1$ but which are not continuous along
the lines $\tilde x = \pi r$, $\tilde y = \pi r$. 

We have not succeeded in finding real derivations of $M_n$ which tend
to real smooth vector fields on the tours. The limit
$n\rightarrow\infty$ is a rather singular limit and it need not be true
that an arbitrary vector field on the torus is the limit of a
derivation.  We constructed the algebra $M_n$ using generators and
relations. This is the noncommutative version of the method of defining a
curved manifold by an embedding in a higher-dimensional flat euclidean
space.  This procedure works well for the sphere but the flat torus
possesses no such embedding. We refer to the book by Thorpe~\cite{Tho79}
for a discussion of this point.

The 1-forms dual to the derivations (3.2.16) are given by
$$
\theta^1 = - ir \epsilon_1 (1 - {1 \over 2} P_1) u^{-1} du, \qquad
\theta^2 = - ir \epsilon_2 (1 - {1 \over 2} P_2) v^{-1} dv.     \eqno(3.2.18)
$$
These are almost as singular as the limit of the expressions given by
(3.2.10). It is important however for us to have the derivations to
define the Laplace operator.

We define the action $S_{1,n}(f)$ of the matrix $f$ on the surface
$\Sigma_1$ as the trace
$$
S_{1,n}(f) = 
{1 \over n} \tr (f^* (\Delta_1 f + \mu^2) f + V(f^*f))           \eqno(3.2.19) 
$$
where the laplacian is the covariant laplacian with respect to the
geometry we have put on the torus and $V(f^*f)$ is an arbitrary
(positive) potential function. The normalization has been chosen so that
$$
\lim_{n \rightarrow \infty} S_{1,n}(f) =  \tilde S_1(\tilde f)   \eqno(3.2.20)
$$
where $\tilde S_1(\tilde f)$ is the usual action of the classical complex 
scalar field $\tilde f$. From (3.2.16) we find
$$
{1 \over n} \tr (u^* \Delta_1 u) = 
{1 \over n} \tr (e_\alpha u^* e^\alpha u) = 
{1 \over n r^2} \tr (u^* u) = (1 + {1\over n}) {1\over r^2}
$$
and similarly for $v$. 

Obviously we shall have
$$
S_{1,n}(f) = 0(n)
$$
for almost all elements $f \in M_n$.  As an example consider the
`coordinate' $x^3$ on the sphere. With the conventions we
have been using one finds the expression
$$
x^3 = {2r \over n} \sum_1^{n-1} {u^l \over 1 - q^{-l}}.         \eqno(3.2.21)
$$
The numerical factor in this expression is valid only for large values
of $n$. Since 
$$
\tr (e_1 u^{*l} e_1 u^l) = o(l^2)
$$
there follows then the estimate
$$
S_{1,n}(x^3) = o(n).                                            \eqno(3.2.22)
$$
At least one of the `coordinates' of the fuzzy sphere becomes singular
then when considered as an element of the fuzzy torus.

\subsection{Higher genera}

An introduction to general Riemann surfaces can be found for example in
the lecture notes by Schlichenmaier~\cite{Sch89}. The algebra of
functions on each surface has been `quantized' using general
$C^*$-algebras~\cite{KliLes92a,KliLes92b}. We conjecture in fact that
this can be done using matrix algebras and that differential calculi can
be constructed over $M_n$ which tend in some way to the de~Rham
differential calculus of $\Sigma_h$ for each genus $h$. The 
construction of Berezin~\cite{Ber75} as well as the fact that
each $\Sigma_h$ can be endowed with a metric of constant Gaussian
curvature is some encouragement. If the differential calculus is based
on derivations then one can define a Laplace operator $\Delta_h$ and an
action
$$
S_{h,n}(f) =
{1 \over n} \tr (f^* (\Delta_h f + \mu^2) f + V(f^*f))            \eqno(3.3.1)
$$
with
$$
\lim_{n \rightarrow \infty} S_{h,n}(f) =  \tilde S_h(\tilde f)
$$
where $\tilde S_h(\tilde f)$ is the usual action of the classical complex 
scalar field $\tilde f$ on the Riemann surface $\Sigma_h$.

\subsection{Smooth transitions}

A generic classical field $\tilde f$ on a surface $\Sigma_h$ of genus
$h$ must have finite action $\tilde S_h(\tilde f)$ and every other
action $\tilde S_{h^\prime}(\tilde f)$ must be infinite. If during the
time evolution the action changes so that $\tilde f$ has finite action
for some other genus $h^\prime$ then this means that the surface has
evolved towards a different topology.  To describe a topological
transition from the sphere to the torus one introduces a `temperature'
$\beta$ and an action $\tilde S_{h(\beta)}$ such that $h(\beta) = 0$ for
$\beta < \beta_c$ and $h(\beta) = 1$ for $\beta > \beta_c$.  The
transition will be of first order. It can be made to be of infinite
order by choosing $h(\beta) = 0$ for $\beta < \beta_c - \epsilon$ and
$h(\beta) = 1$ for $\beta > \beta_c + \epsilon$ and choosing as action 
a smooth functional
$$
\tilde S_{h(\beta)} = ( 1 - p(\beta)) \tilde S_0 + p(\beta) \tilde S_1
$$
in the region $\beta_c - \epsilon \leq \beta \leq \beta_c + \epsilon$.
One task of a noncommutative version of gravity would be to motivate
this {\it ad hoc} change of action functional, to calculate, that is,
the function $p(\beta)$.

The partition function for a complex scalar field over a surface of
genus $h = h(\beta)$ is given by
$$
\tilde Z_{h(\beta)} =
\int e^{- \tilde S_{h(\beta)} [\tilde f]} d \tilde f.           \eqno(3.4.1)
$$
The matrix approximation~\cite{Mad95} is given by
$$
Z_{h(\beta),n} = \int e^{- S_{h(\beta),n} [f]} df               \eqno(3.4.2)
$$
where the path integral is now a well-defined integration over matrices.
We suppose that the `real' value of $n$ is `large' but not infinite,
given by (2.1.1) or (2.2.4). We can then claim that the expression
(3.4.2) is the `correct' one and (3.4.1) is the approximation. For
$\beta < \beta_c$ the contributions from almost all those matrices $f$
which approximate functions on the torus (and other genera) are
suppressed since $S_{1,n} [f] = 0(n)$. On the other hand for
$\beta > \beta_c$ the contributions from almost all those matrices $f$
which approximate functions on the sphere (and other genera) are
suppressed since $S_{0,n} [f] = 0(n)$.

\section{$D$-branes}

Matrices can also be used to give a finite `fuzzy' description of the
space complementary to a Dirichlet $p$-brane, a description which will
allow one perhaps to include the reasonable property that points should
be intrinsically `fuzzy' at the Planck scale.  Strings naturally play a
special role here since they have a world surface of dimension two and
an arbitrary matrix can always be written as a polynomial in two given
matrices. We refer to the literature for a description of Dirichlet
branes in general~\cite{Pol96, BonChu97, Dij97} and within the context
of $M$(atrix)-theory~\cite{BanFisSheSus96, GanRamTay96, HoWu96, Ban97}. 
The action of the matrix description of the complementary space is
conjectured~\cite{deWHopNic88} to be associated to the action in the
infinite-momentum frame of a super-membrane of dimension $p$. Since
quite generally the compactified factors of the surfaces normal to the
$p$-branes are of the Planck scale we conclude from the arguments of the
previous sections that they have ill-defined topology and that a matrix
description will include a sum over many topologies.

We consider a $d$-dimensional manifold $V_d$ with a Kaluza-Klein
reduction to a Dirichlet brane $\Sigma_p$ of dimension $p$. The
$\Sigma_p$ is known as a $(p-1)$-brane. The manifold $V_d$ is therefore
a bundle over $\Sigma_p$ with fibre an $(d-p)$-dimensional manifold
$N_{d-p}$.  We shall suppose for simplicity that the fibration is
trivial, $V_d = \Sigma_p \times N_{d-p}$, and that all manifolds are
parallelizable.  We shall suppose also, as is usual in Kaluza-Klein
theory, that $N_{d-p}$ is space-like.  Let Greek indices $(\alpha,
\beta, \dots)$ take the values 1 to $p$, Latin indices $(a, b, \dots)$
the values $p+1$ to $d$ and Latin indices $(i, j, \dots)$ the values $1$
to $d$. We introduce a moving frame 
$\theta^i = (\theta^\alpha, \theta^a)$ on $V_d$ with $\theta^\alpha$
a moving frame on $\Sigma_p$.  Consider now an electromagnetic field on
$V_d$ and write the field strength $F$ as
$$
F = {1 \over 2} F_{ij} \theta^i \theta^j.
$$ 
Then the electromagnetic action in $V_d$ takes the form
$$
S = {1 \over 4 g^2} \int_{\Sigma_p} 
\int_{N_{d-p}} F_{ij} F^{ij} d^{d-p} x d^p x.
$$

Although we have argued elsewhere~\cite{Mad95} that the entire $V_d$
should be described by a noncommutative algebra we shall suppose here
that the $D$-brane can be described by an ordinary smooth manifold and
that only the $N_{d-p}$ need be `quantized'.  This means that the
algebra ${\cal C}(N_{d-p})$ of (smooth) functions on $N_{d-p}$ is
replaced by a finite-dimensional matrix algebra, the algebra $M_n$ of $n
\times n$ matrices, a procedure which is analogous to the quantization
of a compact phase space~\cite{Ber75}, for example spin.  It means
also that the algebra of de~Rham differential forms $\Omega^*({\cal
C}(N_{d-p}))$ on $N_{d-p}$ must be replaced by a differential calculus
over $M_n$.  This entire procedure has just been described for $d-p = 2$
in Section~4. For $d-p = 4$ we refer to Grosse {\it et
al.}~\cite{GroKliPre97a} and for general $d-p$ to Madore~\cite{Mad96}.
In the case $d-p = 2$ we have seen that only genus zero and one can be
considered with any success.

The components $\theta^a$ of the moving frame introduced on $V_d$ are
to be replaced by a noncommutative equivalent such that in some sense we
have
$$
\lim_{n \rightarrow \infty} \Omega^*(M_n) = 
\Omega^*({\cal C}(N_{d-p})).
$$
We have seen in Section~4 how difficult this limit is to define even for
$p=2$.  We shall use the same symbol to denote a de~Rham form and the
equivalent fuzzy form. Let $\omega$ be an element of 
$\Omega^1({\cal C}(V_d))$. Then in typical Kaluza-Klein fashion we can
write $\omega$ as the sum of a `horizontal' term 
$A = \omega_h$ in $M_n \otimes \Omega^1({\cal C}(\Sigma_p))$ and a
`vertical' term $\omega_v$ in 
${\cal C}(\Sigma_p) \otimes \Omega^1(M_n)$.  More details of this can be
found elsewhere~\cite{Mad95}.

\subsection{Curved complements}

The case in which the space $N_{d-p}$ complementary to a $p$-brane is
curved and compact is the easiest to treat conceptually from a fuzzy
point of view. It contains at least a simple generic example, 
$N_{d-p} = S^2$ which has been worked out in detail and although we
formulate them more generally, most of the following calculations have
been shown to be valid only in this one example.  The case has however
the drawback in that, being curved, the models have no immediate
supersymmetric extension. A rudimentary version of `noncommutative
supergravity' would have to be developed for this purpose.  This would
involve introducing besides the noncommuting bosonic generators a set of
non-anticommuting fermionic generators and defining a linear connection
on the entire structure.  This has yet to be done.

We identify the gauge transformations on $M_n$ as the unitary elements
$U_n$ of $M_n$.  The way in which $U_n$ can be identified with the
local $U_1$ transformations in the commutative limit has been explained 
by Grosse \& Madore~\cite{GroMad92} in the case $p=2$ and genus zero. 
A gauge transformation is therefore given by
$$
\omega \mapsto g^{-1} \omega g + g^{-1} dg                         \eqno(4.1.2)
$$
with $g$ an element of ${\cal U}_n$, the group of local gauge
transformations on $\Sigma_p$. 

Now $\Omega^1(M_n)$ has a preferred 1-form $\theta$ which is
invariant~\cite{DubKerMad90} under a gauge transformation:
$$
\theta \mapsto g^{-1} \theta g + g^{-1} d_v g = \theta.           \eqno(4.1.3)
$$
We have here decomposed $d = d_v + d_h$.  Choose an integer $m$ and an
anti-hermitian basis $\lambda_a$ of the Lie algebra of $SU_m$. Restrict
$n$ to those values such that $SU_m$ has an irreducible representation
of dimension $n$ and restrict the $N_{d-p}$ to be an orbit of the
adjoint representation of $SU_m$. For example if $m=2$ then $n$ can take
any values and $d-p = 2$ which is the dimension of $SU_m$ minus the
number of Casimir operators.  If $m=3$ then $d-p = 6$ which is again the
dimension of $SU_m$ minus the number of Casimir operators. The manifold
$N_{d-p}$ is a manifold embedded in $S^7$ and defined by the cubic
Casimir operator of $SU_3$.  If we introduce the derivations 
$e_a = \ad \lambda_a$ and the 1-forms $\theta_a$ dual to them then
$\theta = - \lambda_a \theta^a$.  (In the limit (4.1.1) $\theta$ is
singular~\cite{Mad92} in the case $p=2$ and genus zero; this is because
the sphere is not parallelizable and in the commutative limit the
$\theta^a$ must be defined on a parallelizable bundle over it) We can
decompose then
$$ 
\omega = A + \theta + \phi                                        \eqno(4.1.4)
$$
where $\phi = \omega_v - \theta$ is the difference between two connections
and so transforms under the adjoint representation of $U_n$.  We write
$\phi = \phi_a \theta^a$. Then a straightforward calculation leads to
the identities
$$
F_{ab} = [\phi_a, \phi_b] - C^c{}_{ab}\,\phi_c, \qquad
F_{\alpha a} = D_\alpha \phi_a                                    \eqno(4.1.5)
$$
for the `fuzzy' and mixed components of the electromagnetic field strength.
The structure constants $C^c{}_{ab}$ are defined with respect to the basis
$\lambda_a$ of $SU_m$. 

A rather dubious mathemetical argument leads, at least in the case
$d-p=2$ and genus zero~\cite{GroMad92}, to the limit
$$
\lim_{n \rightarrow \infty} {1 \over n}  S_n = S.                 \eqno(4.1.6)
$$
The $S_n$ is given by
$$
S_n = {1\over 4g^2}  \int_{\Sigma_p} \tr(F_{ij} F^{ij}) = 
{1\over 4g^2} \int_{\Sigma_p} \tr(F_{\mu\nu} F^{\mu\nu}) - 
{1\over 2g^2} \int_{\Sigma_p} \tr (D_\mu \phi_a) (D^\mu \phi^a) +V(\phi)
                                                                  \eqno(4.1.7)
$$
where 
$$
V(\phi) = - {1\over 4g^2} \int_{\Sigma_p}  \tr(F_{ab} F^{ab})     \eqno(4.1.8)
$$
is the potential. We see that it can be thought of as the field strength 
in the fuzzy directions. 

If we consider the $\phi_a$ as `coordinates' on the fuzzy version of
$N_{d-p}$ then $\Sigma_p$ is given by $\phi_a = 0$, which is a stable
zero of the potential (4.1.8). Another obvious stable zero is given by
$\phi = \lambda_a$. There are in general other stable zeros, the
number of which increases with $n$.  For example, in the case of the
2-sphere there are in all $p(n)$ (the partition function) zeros of
$V(\phi)$. One can think of this as meaning that there are $p(n)$
possible `positions' for the $D$-branes and they are all stable with
the same potential energy. Energy is required however to transit from
one state to another. In a complicated way the number of massless modes
increases as one `approaches' the vacuum $\phi_a = 0$. By this we mean
that when $\phi_a = 0$ there is a $U_n$ multiplet of massless modes,
when $\phi = \lambda_a$ there is only a $U_1$ multiplet and the number
of massless modes in the $p(n) - 2$ vacuua between these two extremes
depends on the characteristics of the vacuum.

The $M_n$ are curved `manifolds' in general and endowed with a linear
connection. The covariant derivatives $D_a \phi_b$ of $\phi_a$ in the
directions normal to $\Sigma_p$ are given by
$$
D_a \phi_b = [\phi_a, \phi_b] - {1 \over 2} C^c{}_{ab} \phi_c =
F_{ab} + {1 \over 2} C^c{}_{ab} \phi_c.                            \eqno(4.1.9)
$$
It vanishes therefore on the stable vacuum given by $\phi_a = 0$ but not
on the others.  Except for the `curvature term' in the expression of the
`vertical' components $F_{ab}$ of the Yang-Mills field strength the
action~(4.1.7) is identical to the bosonic part of the one which has
been proposed in $M$(atrix) theory, To see if it is possible to obtain
exactly the $M$(atrix)-theory action we turn our attention to flat
complements $N_{d-p}$.

\subsection{Flat complements}

The simplest example of a compact manifold $N_{d-p}$ which could admit a
flat metric is the $(d-p)$-torus. We have not succeeded in treating this
case for general values of $d-p$ but it would seem from our
considerations of Section~3.2 that it is very difficult if not
impossible to define a differential calculus on a noncommutative version
of the 2-torus which tends smoothly to the de~Rham differential calculus
and which admits a flat metric. The case of a flat $N_{d-p}$ is
paradoxally more difficult to treat from a fuzzy point of view than the
curved one.

\section*{Acknowledgment} The authors would like to thank A. Kehagias 
and S. Theisen for enlightening conversations.  One of the authors (JM)
would like to thank J. Wess for his hospitality at the Ludwig-Maxmillian
Universit\"at, M\"unchen and the other (LAS) would like to thank CNPq,
Brazil for financial support.

\end{document}